\documentclass[reprint,superscriptaddress,longbibliography]
      {revtex4-1}

\usepackage{siunitx}
\usepackage{bm}
\usepackage{xcolor}
\usepackage{graphicx}
\usepackage[utf8]{inputenc}
\usepackage{epstopdf}
\usepackage{afterpage}
\usepackage{setspace}
\usepackage{booktabs}
\usepackage{hyperref}
\usepackage{subcaption}
\usepackage{amsmath}
\usepackage{scalerel}
\usepackage{float}

\expandafter\ifx\csname package@font\endcsname\relax\else
 \expandafter\expandafter
 \expandafter\usepackage
 \expandafter\expandafter
 \expandafter{\csname package@font\endcsname}%
\fi
\begin{document}

\setstretch{1.0}

\title{Machine learning potentials for redox chemistry in solution}

\author{Emir Kocer}
\affiliation{Lehrstuhl f\"ur Theoretische Chemie II, Ruhr-Universit\"at Bochum, 44780 Bochum, Germany} 
\affiliation{Research Center Chemical Sciences and Sustainability, Research Alliance Ruhr, 44780 Bochum, Germany}
\author{Redouan El Haouari}
\affiliation{Lehrstuhl f\"ur Theoretische Chemie II, Ruhr-Universit\"at Bochum, 44780 Bochum, Germany} 
\affiliation{Research Center Chemical Sciences and Sustainability, Research Alliance Ruhr, 44780 Bochum, Germany}
\author{Christoph Dellago}
\affiliation{University of Vienna, Faculty of Physics, Boltzmanngasse 5, A-1090 Vienna, Austria}
\author{J\"{o}rg Behler}
\email{joerg.behler@ruhr-uni-bochum.de}
\affiliation{Lehrstuhl f\"ur Theoretische Chemie II, Ruhr-Universit\"at Bochum, 44780 Bochum, Germany} 
\affiliation{Research Center Chemical Sciences and Sustainability, Research Alliance Ruhr, 44780 Bochum, Germany}

\begin{abstract}
Machine learning potentials (MLPs) represent atomic interactions with quantum mechanical accuracy offering an efficient tool for atomistic simulations in many fields of science. However, most MLPs rely on local atomic energies without information about the global composition of the system. To date, this has prevented the application of MLPs to redox reactions in solution, which involve chemical species in different oxidation states and electron transfer between them.
Here, we show that fourth-generation MLPs overcome this limitation and can provide a physically correct description of redox chemical reactions. For the example of ferrous (Fe$^{2+}$) and ferric (Fe$^{3+}$) ions in water we show that the correct oxidation states are obtained matching the number of chloride counter ions irrespective of their positions in the system. 
Moreover, we demonstrate that our method can describe electron-transfer processes between ferrous and ferric ions, paving the way to simulations of general redox chemistry in solution.
\end{abstract}

\maketitle 

\section{Introduction}\label{sec:introduction}

Redox reactions, in which electrons are transferred from one chemical species to another, play a fundamental role across many fields of chemistry. Important examples are photosynthesis and enzymatic reactions that drive the processes of life, electrochemical water splitting for green hydrogen production, energy storage and conversion in batteries and fuel cells, as well as the corrosion of materials. Moreover, redox chemistry is considered as pivotal for the electrification of the chemical industry through transforming processes from fossil fuels to more sustainable renewable energy sources -- an essential shift for maintaining the standard of living in modern societies~\cite{P6694}. 

Due to the importance of redox reactions, a substantial effort has been devoted to understanding their underlying mechanisms in detail, and in recent years computer simulations have increasingly contributed to these efforts~\cite{P6839,P6840}. 
To date, most of these simulations rely on accurate but computationally demanding quantum mechanical electronic structure methods, such as density-functional theory (DFT), to compute the atomic interactions. As a result, the length and time scales accessible in these \emph{ab initio} molecular dynamics (AIMD) simulations remain limited, restricting their application to relatively simple model systems. More efficient potentials such as classical force fields~\cite{P6825,P6827}, which have been successfully employed in large-scale simulations in other fields of chemistry, are usually unable to describe electron transfer processes and the change of atomic oxidation states during the simulation. While a few advanced reactive force fields can overcome this limitation~\cite{P5669},
they usually fall short of reaching ``chemical accuracy'' and thus lack the predictive power of electronic structure methods. For these reasons, theoretical studies of complex redox reactions in solution have remained a significant challenge and are essentially confined to the domain of quantum chemistry.

In recent years, rapid advances in machine learning techniques have driven a paradigm change in the construction of atomistic potentials. Rather than relying on the cumbersome development of increasingly sophisticated yet inherently approximate physical models, modern machine learning potentials (MLP)~\cite{P4885,P6102,P5673,P6121,P6112} ``learn'' the high-dimensional potential energy surface (PES) -- which encapsulates all information about atomic interactions -- directly from high-level reference electronic structure data. 
Consequently, MLPs allow to compute the energy and forces as a function of the atomic positions with an accuracy approaching that of quantum mechanics, while maintaining efficiency comparable to that of simple empirical force fields.

Although numerous MLPs have been developed for a wide range of systems, they have not yet been applicable to redox reactions in solution. This obstacle arises because redox reactions involve chemical species in multiple oxidation states, each of which interacts differently with its environment. MLPs can capture redox reactions only in some cases, for instance if a metal surface is oxidized to form a surface oxide. In such cases, the coordination of the metal atoms by neighboring oxide ions acts as a structural label for the oxidation state. Such a label then enables the construction of MLPs where the atomic energy contributions solely depend on the local environments. Even relatively small geometric changes like Jahn-Teller distorted environments can be sufficient for this purpose~\cite{P5867}. Unfortunately, the situation is completely different in a liquid environment, where mobile solvent molecules and other chemical species in solution do not provide a fixed matrix that is suitable as label to define the oxidation state. Thus, in liquid solvents, i.e., where chemical reactions typically take place, most MLPs face significant challenges in distinguishing between different oxidation states. Consequently, even for simple ions like multivalent transition metal cations it is impossible to predict the oxidation state and thus accurate interactions with other species in the system.

Since the advent of MLPs about thirty years ago~\cite{P0316}, significant progress has been made in extending their applicability to increasingly complex systems. While early first-generation MLPs were restricted to very small systems consisting of only a few atoms, second-generation MLPs extended the applicability to large condensed systems containing thousands of atoms. This advance is achieved by constructing the total energy as a sum of atomic energies that depend on the local chemical environment defined by a cutoff radius~\cite{P1174}. Many types of second-generation MLPs have been introduced to date and successfully applied to many problems in chemistry and materials science~\cite{P1174,P2630,P4945,P4862,P5794}. Long-range electrostatic interactions based on environment-dependent atomic charges are included in third-generation MLPs~\cite{P2391,P2962,P5577,P5885}. 

Due to the approximation of local atomic energies and charges in second- and third-generation MLPs, they lack important information about the system's overall chemical composition and the resulting global charge distribution, both of which are necessary for determining atomic oxidation states. In contrast, fourth-generation MLPs, which consider the charge distribution in the entire system, e.g., by charge equilibration~\cite{P1448,P4419,P5932,P6666,P6829} or some form of iterative procedure~\cite{P5859,P6122}, in principle contain this information. This raises the question of whether, and to what extent, these models are able to accurately describe redox chemistry in solution.

In this work, we assesses the capabilities of different generations of MLPs in describing redox reactions in solution, using the family of high-dimensional neural network potentials (HDNNP)~\cite{P6018}. Specifically, we use 2G-HDNNPs~\cite{P1174} and 4G-HDNNPs~\cite{P5932} as typical examples for local second- and global fourth-generation MLPs, respectively. As benchmark systems we have chosen FeCl$_2$ and FeCl$_3$ in liquid water. These systems are very challenging for MLPs since the oxidation states of ferrous (Fe$^{2+}$) and ferric (Fe$^{3+}$) ions, which are both stable in water, are determined by the number of chloride counter ions in the system. If at least some of these anions are outside the local iron atomic environments, the oxidation states are ambiguous and it is anticipated that second-generation potentials cannot distinguish between the two states. In particular if a MLP is trained to a combined dataset of solvated FeCl$_2$ and FeCl$_3$, the local chemical compositions of the iron atomic environments can be similar for both ions, while the DFT interatomic forces are clearly different. In such a situation, a 2G-HDNNP will learn an averaged interaction and thus both ions will interact in the same way with the solvent, leading to unphysical and qualitatively wrong solvation spheres. Fourth-generation MLPs, however, should in principle be able to describe both oxidation states, Fe$^{2+}$ and Fe$^{3+}$, correctly, since they take global information about the system into account. By constructing 2G- and 4G-HDNNPs for different datasets, we will investigate these hypotheses and evaluate the accuracy of the resulting potentials in a realistic application. Moreover, molecular dynamics (MD) simulations will be used to determine the structure of the solvation spheres in each case to identify possible shortcomings of the potentials.

Finally, we extend our analysis beyond the isolated FeCl$_2$ and FeCl$_3$ systems and perform MD simulations of a large periodic box containing both, one FeCl$_2$ as well as one FeCl$_3$ formula unit, in liquid water. This Fe$_2$Cl$_5$ system will enable us to study possible electron transfer processes between a ferrous iron ion and the ferric iron ion, representing one of the most prominent and most frequently studied model systems for electron transfer reactions in aqueous solution \cite{P0871,P3687,P6457,P6802,P6811,P6812,P6813,P6831,P6832}.

\section{Results}\label{sec:results}

\subsection{Neural Network Potentials}\label{sec:potentials}

\begin{figure}[!]
    \centering
    \includegraphics[width=0.45\textwidth]{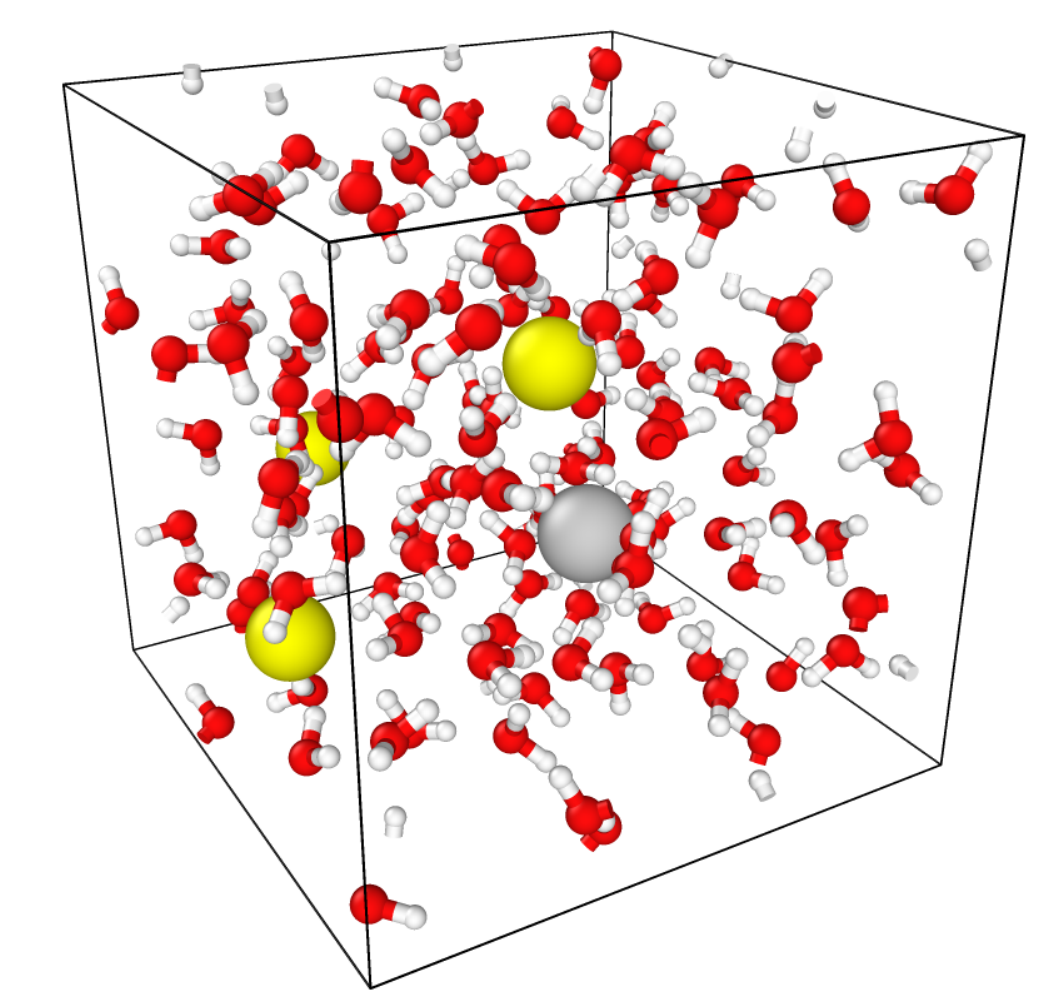}
    \caption{Structure of the medium-sized box of the FeCl$_3$ system. The grey sphere represents the iron atom, the three yellow spheres are the chlorine atoms.}
    \label{fig:box}
\end{figure}

For investigating the applicability of MLPs to redox reactions in solution we construct a series of  FeCl$_2$ and FeCl$_3$ systems in liquid water. These consist of periodic cubic boxes of three different sizes called ``small'', ``medium'', and ``large'', with side lengths of approximately 10, 15, and 30~\AA{}, respectively. An example of the medium box is given in Fig.~\ref{fig:box}. They are filled with water at ambient conditions as well as a single formula unit of either FeCl$_2$ or FeCl$_3$, resulting in systems containing roughly 100, 380 and 2970 atoms. 

The small and medium-sized boxes have been used in DFT calculations to generate the reference energies and forces, and in case of 4G-HDNNPs also Hirshfeld charges~\cite{P0416}, for training the potentials. On average, these charges have numerical values of 0.35$e$ for Fe$^{2+}$ and of 0.52$e$ for Fe$^{3+}$. This corresponds to the anticipated charge ratio of 2:3 and thus allows to identify both oxidation states. Other alternative charge partitioning schemes would provide different numerical values but qualitatively the same description and thus could be equally used. The medium and large boxes will be used in HDNNP-driven MD simulations to study the oxidation states of the iron ions and their solvation spheres.  

Employing the DFT datasets that are described in detail in the SI, six different HDNNPs have been constructed with the aim to assess the ability of 2G- and 4G-HDNNPs to describe the different types of datasets and oxidation states. The HDNNPs labeled 2G(Fe$^{2+}$), 2G(Fe$^{3+}$), and 2G(Fe$^{2+}$/Fe$^{3+}$) employ second-generation HDNNPs and have been trained only to the FeCl$_2$ data, only to the FeCl$_3$ data or to the combined full dataset of both systems, respectively. Accordingly, the HDNNPs labeled 4G(Fe$^{2+}$), 4G(Fe$^{3+}$), and 4G(Fe$^{2+}$/Fe$^{3+}$) employ fourth-generation HDNNPs for the same datasets. 

The 2G(Fe$^{2+}$) and 2G(Fe$^{3+}$)-HDNNPs are expected to perform well for the respective isolated oxidation state since there is only one ``type'' of iron interacting with water resulting in a unique equilibrium solvent structure for each case. Still, in general, local second-generation potentials do not have sufficient information about the number of chloride ions present in the system as these might be outside the local Fe atomic environments. Thus 2G-HDNNPs such as 2G(Fe$^{2+}$/Fe$^{3+}$) trained to the combined FeCl$_2$/FeCl$_3$ dataset, i.e., two different solvation structures around iron, cannot uniquely determine the oxidation state-dependent interactions between Fe and the surrounding water molecules and are anticipated to result in potentials of poor quality. 4G-HDNNPs, on the other hand, make use of global information and are expected to be able to determine the iron oxidation state and the resulting solvent structure correctly even when trained to the combined dataset containing Fe$^{2+}$ and  Fe$^{3+}$.

Table~\ref{tab:RMSE_table} shows the root mean square errors (RMSE) of the energies, forces and atomic charges for the training sets and the independent test sets of all six potentials. Interestingly, for all datasets the energy and force RMSEs of all potentials, 2G-HDNNPs as well as 4G-HDNNPs, are very low and of the typical order of magnitude expected for state-of-the-art MLPs. Therefore, at first glance, all HDNNPs seem to provide a similarly accurate description of all datasets and systems, despite the conceptual limitations of 2G-HDNNPs. Hence, further tests beyond simple RMSE values are clearly required to identify possible shortcomings of the potentials.

\begin{table*}[t] 
\centering
\caption{Root mean square errors (RMSE) for the energies per atom, atomic force components and atomic partial charges
of the training sets (test set values in brackets) for all potentials developed in this work. 
}
\label{tab:RMSE_table}
{\footnotesize 
\begin{tabular}{@{}l|c|c|c|c|c|c@{}} 
\toprule
HDNNP &2G(Fe$^{2+}$) & 2G(Fe$^{3+}$) & 2G(Fe$^{2+}$/Fe$^{3+}$) & 4G(Fe$^{2+}$) & 4G(Fe$^{3+}$)& 4G(Fe$^{2+}$/Fe$^{3+}$)  \\
\midrule
\hline
$E$ (meV/atom) & 0.203 (0.213) & 0.232 (0.254) & 0.262 (0.271) & 0.203 (0.212) & 0.237 (0.258) & 0.258 (0.273)\\
$F$ (eV/Bohr) & 0.032 (0.032) & 0.037  (0.037) & 0.034 (0.035) & 0.031 (0.031) & 0.036 (0.036) & 0.033 (0.033)\\
$Q$ (me) & - & - & - & 2.896 (2.896) & 3.130 (3.124) & 3.105 (3.108) \\
\bottomrule
\end{tabular}
} 
\end{table*}

\subsection{2G-HDNNP Simulations}\label{sec:2GHDNNP}

\begin{figure}[!]
    \centering
    \includegraphics[width=0.45\textwidth]{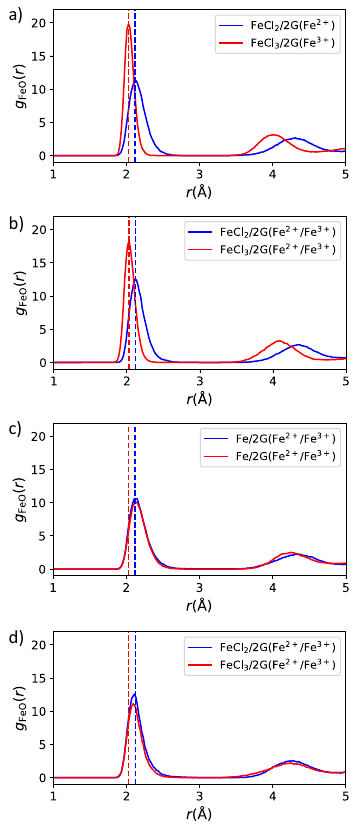}
    \caption{Fe-O radial distribution functions of Fe$^{2+}$ and Fe$^{3+}$ in water obtained with 2G-HDNNP potentials. The RDFs in (a) have been obtained for the medium box from 2G-HDNNPs trained to the separate FeCl$_2$ and FeCl$_3$ datasets, respectively, and the vertical lines define our \emph{ab initio} reference first peak positions for the two iron oxidation states. Panel (b) shows the same RDFs obtained as well for the medium box but using a 2G-HDNNP trained to the combined FeCl$_2$ and FeCl$_3$ datasets. Panel (c) shows the RDFs obtained for both ions using the 2G-HDNNP trained to the combined FeCl$_2$ and FeCl$_3$ datasets after removing all chlorine atoms in the system. The RDFs in (d) have been computed starting from FeCl$_2$ and FeCl$_3$ equilibrated in large boxes using the 2G-HDNNP trained to the combined FeCl$_2$ and FeCl$_3$ datasets.
} 
    \label{fig:2G-RDFs}
\end{figure}

As a first test, we have performed MD simulations of the FeCl$_2$ system using HDNNP 2G(Fe$^{2+}$) and of the FeCl$_3$ system using HDNNP 2G(Fe$^{3+}$). Since these potentials have been trained on data for a single oxidation state only, even as local potentials they are expected to accurately describe the solvent structure of the respective iron ion. The Fe-O radial distribution functions (RDF) from 100~ps trajectories in the medium box size are shown in Fig.~\ref{fig:2G-RDFs}a. The first peak for Fe$^{3+}$-O is located at a relatively short distance of about 2.03~\AA{} due to the strong interaction between the highly charged Fe$^{3+}$ cation and the six coordinating water molecules, while for the weaker interaction between Fe$^{2+}$ and water the first Fe$^{2+}$-O peak position is found at a slightly larger distance of 2.12~\AA{}. Another distinction between the two Fe-O RDFs is the higher amplitude and narrower distribution in case of Fe$^{3+}$, which can also be attributed to the stronger attraction between Fe$^{3+}$ and the surrounding water molecules. Given the uncertainty of the employed exchange-correlation functional, this is in very good agreement with previous work, in which the Fe-O distance of Fe$^{2+}$ in water is typically around 2.0-2.2 \AA{} \cite{P6820,P6457}, while for Fe$^{3+}$ it is more commonly around 1.9-2.05 \AA{} \cite{P6823,P6830}. 
Hence, these two peak positions will from now on be defined as our \emph{ab initio} reference positions for Fe$^{2+}$ and Fe$^{3+}$ and will be highlighted as vertical dashed lines in all RDF plots.

Next, we performed MD simulations of the FeCl$_2$ and FeCl$_3$ systems employing HDNNP 2G(Fe$^{2+}$/Fe$^{3+}$). This HDNNP has been trained to data for both oxidation states and thus in principle should not be able to discriminate correctly the different solvent structures around both Fe ions.
For the initial atomic configurations we made sure that some chloride ions are outside the atomic environment cutoff radius of the Fe ions. This way the Fe ions do not have direct information about their oxidation states by ``counting'' the choride ions based on information in the descriptor values characterizing the geometric atomic environments. Still, the results for the medium box size shown in Fig.~\ref{fig:2G-RDFs}b are essentially indistinguishable from the RDFs in Fig.~\ref{fig:2G-RDFs}a and seem to be physically correct for both oxidation states. 
A closer analysis shows that the reason for this ability of HDNNP 2G(Fe$^{2+}$/Fe$^{3+}$) to distinguish both systems is the still rather moderate distance between the Fe ions and the Cl ions in the medium-sized system. As confirmed in the Fe-Cl distance plots along both trajectories shown in Fig.~5 in the SI the chlorine atoms are outside the Fe cutoff spheres most of the time. However, there are ``bridging'' water molecules which are still relatively close to both ion types and include Fe as well as Cl atoms inside their cutoff spheres. As a consequence, they are able to form physically correct iron solvation spheres.
A detailed explanation of the origin of such a transport of information over a distance of up twice the cutoff radius can be found in Refs.~\citenum{P5128,P6527}. 

To further test this hypothesis, we next conducted MD simulations for artificial systems of infinitely diluted electrolytes  
generated by
removing all chlorine atoms from both systems in the medium box. This is technically possible because 
2G-HDNNPs neither include explicit electrostatic interactions nor atomic charges, so the removal of chlorine does not result in a charged system. On the other hand, the iron atoms keep interacting with water like ions as learned from the training data. 
As shown in Fig.~\ref{fig:2G-RDFs}c, the chlorine removal causes the first peak positions in both RDFs to shift to the same Fe-O distance  although the simulations started from the different equilibrated Fe$^{2+}$ and Fe$^{3+}$ solvation spheres, respectively. Interestingly, the peak position slightly beyond the reference Fe$^{2+}$-O peak corresponds to the trend in peak position with respect to the amount of chloride in the system. While the oxidation states in these simulations are in principle ill-defined due to the absence of any chloride ions, these results provide clear evidence that the apparently correct solvation in Fig.~\ref{fig:2G-RDFs}b is indeed caused by the chloride ions in the systems. 

Finally, to confirm this finding for systems with chemically meaningful compositions, i.e., with the correct numbers of counter ions, we have repeated the simulations of FeCl$_2$ and FeCl$_3$ using HDNNP 2G(Fe$^{2+}$/Fe$^{3+}$) in a large box for 50~ps, ensuring that the initial separations between Fe and the Cl ions were at least twice the atomic environment cutoff (cf. Fig.~5c,d in the SI). Like in the artificial case without any chlorine atoms, the Fe-O RDFs now exhibit wrong first peaks at the same position for both systems (Fig.~\ref{fig:2G-RDFs}d). This clearly demonstrates that MD simulations employing HDNNP 2G(Fe$^{2+}$/Fe$^{3+}$) are indeed unable to provide a physically correct description of the solvation spheres of both, Fe$^{2+}$ and Fe$^{3+}$, if the chloride ions are outside the close environment. 

\subsection{4G-HDNNP Simulations}\label{sec:4GHDNNP}

\begin{figure}[!]
    \centering
    \includegraphics[width=0.45\textwidth]{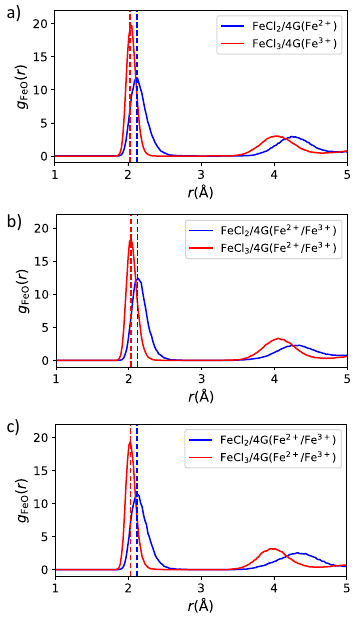}
    \caption{Fe-O radial distribution functions of Fe$^{2+}$ and Fe$^{3+}$ in water obtained with 4G-HDNNP potentials. The RDFs in (a) have been obtained for the medium box from 4G-HDNNPs trained to the separate FeCl$_2$ and FeCl$_3$ datasets, respectively. Panel (b) shows the same RDFs also obtained for the medium box  but using a 4G-HDNNP trained to the combined FeCl$_2$ and FeCl$_3$ datasets. The RDFs in (c) have been computed starting from FeCl$_2$ and FeCl$_3$ equilibrated in large boxes using the 4G-HDNNP trained to the combined FeCl$_2$ and FeCl$_3$ datasets.} 
    \label{fig:4G-RDFs}
\end{figure}

Having demonstrated that second-generation MLPs are generally unable to describe ions in different oxidation states in solution, these simulations are now repeated using fourth-generation HDNNPs. Figure~\ref{fig:4G-RDFs}a shows the Fe-O RDFs of the FeCl$_2$ and FeCl$_3$ systems in the medium-sized box employing HDNNP 4G(Fe$^{2+}$) and HDNNP 4G(Fe$^{3+}$), respectively. Like in the 2G-HDNNP case, also the 4G-HDNNPs trained to the separate FeCl$_2$ and FeCl$_3$ datasets yield the correct solvent structure. Moreover,  the Fe-O RDFs in the medium-sized box are correct for both oxidation states employing HDNNP 4G(Fe$^{2+}$/Fe$^{3+}$) trained to the combined dataset.
However, the critical tests are the large boxes for which the 2G-HDNNP failed (Fig.~\ref{fig:2G-RDFs}d). Even in this case the 4G-HDNNP first peaks obtained for both systems in a 50~ps trajectory are now at the right positions (Fig.~\ref{fig:4G-RDFs}c). Since HDNNP 4G(Fe$^{2+}$/Fe$^{3+}$) produces physically correct solvation structures also for large Fe-Cl separations, we conclude that the non-local information incorporated in 4G-HDNNPs allows to correctly assign the Fe oxidation state irrespective of the distances between the Fe and Cl ions in the system (cf. Fig.~6 in the SI). 

\begin{figure}[!]
    \centering
    \includegraphics[width=0.45\textwidth]{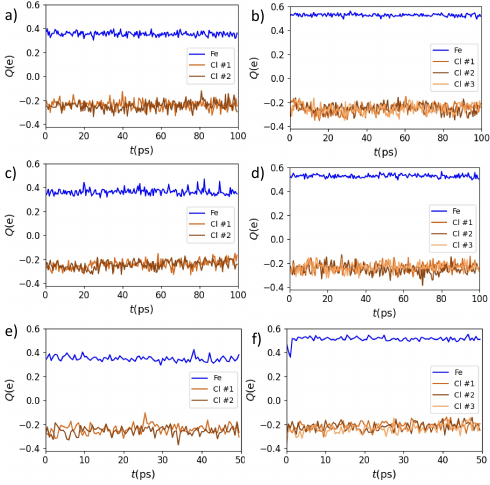}
    \caption{Fe and Cl atomic partial charges predicted by 4G(Fe$^{2+}$) in a medium-sized FeCl$_2$ system (a), by 4G(Fe$^{3+}$) in a medium-sized FeCl$_3$ system (b), by 4G(Fe$^{2+}$/Fe$^{3+}$) in a medium-sized FeCl$_2$ system (c), by 4G(Fe$^{2+}$/Fe$^{3+}$) in a medium-sized FeCl$_3$ system (d), by 4G(Fe$^{2+}$/Fe$^{3+}$) in a large FeCl$_2$ system (e), and by 4G(Fe$^{2+}$/Fe$^{3+}$) in a large FeCl$_3$ system (f).} 
    \label{fig:4G-charges}
\end{figure}

The positions of the first RDF peaks are a convincing but indirect evidence for the correct assignment of the Fe oxidation states. As a further test  we show in Fig.~\ref{fig:4G-charges} the charges of all Fe and Cl ions predicted by the 4G-HDNNPs along the trajectories of Fig.~\ref{fig:4G-RDFs}. In all simulations and for all 4G-HDNNPs, the Fe charges of the FeCl$_2$ system correspond to the reference Hirshfeld charges of Fe$^{2+}$ and the Fe charges of the FeCl$_3$ are in the typical range of the Hirshfeld charges of Fe$^{3+}$. This is a clear evidence that the charge equilibration in the 4G-HDNNP method yields the same charge distributions as the underlying DFT reference method for all systems. These charges, which are a crucial ingredient of the 4G-HDNNP,  ensure the correct description of the Fe-water interactions not only via the electrostatic interactions, which are in fact assumed to play a minor role beyond the rather large cutoff employed here, but in particular as additional part of the input feature vector of the 4G-HDNNP atomic energy neural networks.

Finally, as a side remark we note that the iron and chlorine charges in Fig.~\ref{fig:4G-charges} do not perfectly add to zero as could be expected. The reason for this observation is the underlying Hirshfeld partitioning of the charges in the system. In this scheme, atomic charge densities are overlapping in space such that also solvent molecules close to the ions are formally assigned a small part of the respective ionic charges. This does not have any consequences for our findings, and the overall system charges are exactly zero in all cases due to the total charge constraint applied in the 4G-HDNNP charge equilibration step.

\subsection{Solvent Rearrangement}\label{sec:transferability}

Thus far, we have shown that 4G-HDNNPs are able to describe different oxidation states of an ion in MD simulations. However, the MD simulations of the FeCl$_2$ and FeCl$_3$ systems discussed in the previous sections have been started using equilibrated solvation spheres of the respective ions, which are structurally different and thus might bias the simulation outcome. For instance, it is well-known that different oxidation states, e.g. of transition metal ions, in solids can be distinguished even by 2G-HDNNPs if the local geometric environments are different. An example is the presence or absence of Jahn-Teller distortions in the octahedral environments of Mn$^{3+}$ and Mn$^{4+}$ ions in LiMn$_2$O$_4$~\cite{P5867}. The structurally different solvation spheres of Fe$^{2+}$ and Fe$^{3+}$ might, in principle, play a similar role as the local environments determine the atomic electronegativities of the 4G-HDNNP and therefore may bias the outcome of the charge equilibration.

A hint that the initial solvent structure does not precondition the simulation outcome has been found already in the test trajectories without any chloride ions (cf. Fig.~\ref{fig:2G-RDFs}c), where different initial solvent structures relaxed to the same radial distribution of solvent molecules using a second-generation HDNNP. Moreover, the ability of 4G-HDNNPs to describe different oxidation states is not only rooted in the local geometric environments but also based on global information about the charge distribution and chemical composition of the system. 
Therefore, 4G-HDNNPs offer a much more general approach for difficult systems than 2G-HDNNPs, e.g., in the previously investigated case of LiMn$_2$O$_4$~\cite{P5866}.

\begin{figure}[!]
    \centering
    \includegraphics[width=0.45\textwidth]{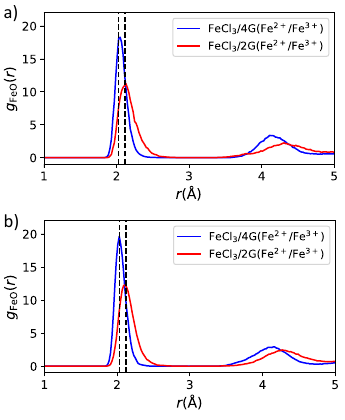}
    \caption{Fe-O RDFs sampled from MD trajectories that have been generated with HDNNP 2G(Fe$^{2+}$/Fe$^{3+}$) and HDNNP 4G(Fe$^{2+}$/Fe$^{3+}$) a) for the medium-sized box and b) for the large-sized box. The initial geometries of the FeCl$_3$ systems have been generated from an equilibrated FeCl$_2$ system by replacing a water molecule at large distance from the Fe ion by a chloride ion. While in the 2G-HDNNP simulation the solvent structure remains that of Fe$^{2+}$, in the 4G-HDNNP simulation the correct solvation structure of Fe$^{3+}$ is obtained.} 
    \label{fig:addition}
\end{figure}

To demonstrate that formation of the correct solvation sphere of Fe$^{2+}$ and Fe$^{3+}$ is indeed independent of the initial structure, we have performed simulations starting from equilibrated solvent structures of FeCl$_2$, with one distant water molecule exchanged by a chlorine atom in a medium-sized box. Consequently, the Fe$^{2+}$ ion and its solvation sphere should switch to Fe$^{3+}$. This test has been performed for both potentials trained to the combined dataset, HDNNP 2G(Fe$^{2+}$/Fe$^{3+}$) and HDNNP 4G(Fe$^{2+}$/Fe$^{3+}$). As can be seen in the Fe-O RDFs in Fig.~\ref{fig:addition}a,  the solvation sphere does not change for HDNNP 2G(Fe$^{2+}$/Fe$^{3+}$) and remains similar to Fe$^{2+}$, which is physically incorrect. For HDNNP 4G(Fe$^{2+}$/Fe$^{3+}$), however, the system is correctly converted to the solvent structure of Fe$^{3+}$. We repeated these simulations employing large boxes to ensure that no interactions between Fe and Cl are involved in these results and found the same failure of the 2G-HDNNP while the 4G-HDNNP again correctly switched the system to Fe$^{3+}$ (Fig.~\ref{fig:addition}b).

\subsection{Transferability: Electron Transfer Reactions}\label{sec:redox}

Since the 4G(Fe$^{2+}$/Fe$^{3+}$) HDNNP is able to describe Fe$^{2+}$ as well as Fe$^{3+}$ ions in aqueous solution, we finally tested the transferability of this potential to water boxes containing two Fe ions and five chloride ions. In this Fe$_2$Cl$_5$ system the electrons should distribute such that one Fe$^{2+}$ and one Fe$^{3+}$ ion are formed. 
In principle, electron transfers between these two ions, which are induced by thermal fluctuations in the structure of the solvation spheres, can occur during MD simulations, which is the basis of Marcus theory of electron transfer~\cite{P3548}. 

We note that here such simulations can provide at most qualitative results for the present parameterization of the 4G(Fe$^{2+}$/Fe$^{3+}$) HDNNP, as it has been trained on systems containing only one Fe atom and at most three Cl atoms. In MD simulations of the Fe$_2$Cl$_5$ system, however, atomic interactions may occur that are not present in the reference training data, such as interactions between two iron ions or water molecules interacting with more than one Fe ion and/or with more than three chloride ions. This extrapolation in chemical composition can be very challenging for MLPs, in particular if new interactions are introduced. Here, the short-range interactions between two iron ions have not been considered at all in the construction of the potential since in the reference structures an iron atom could not be located inside the cutoff sphere of another iron atom due to the applied periodic boundary conditions and box parameters. Consequently, no environment descriptors for Fe-Fe interactions have been included in the HDNNP, and the potential is expected to be unreliable if such interactions become important. In future work this limitation can easily be overcome by extending the reference dataset to include structures with more ions of each species. 

\begin{figure}[!]
    \centering
    \includegraphics[width=0.45\textwidth]{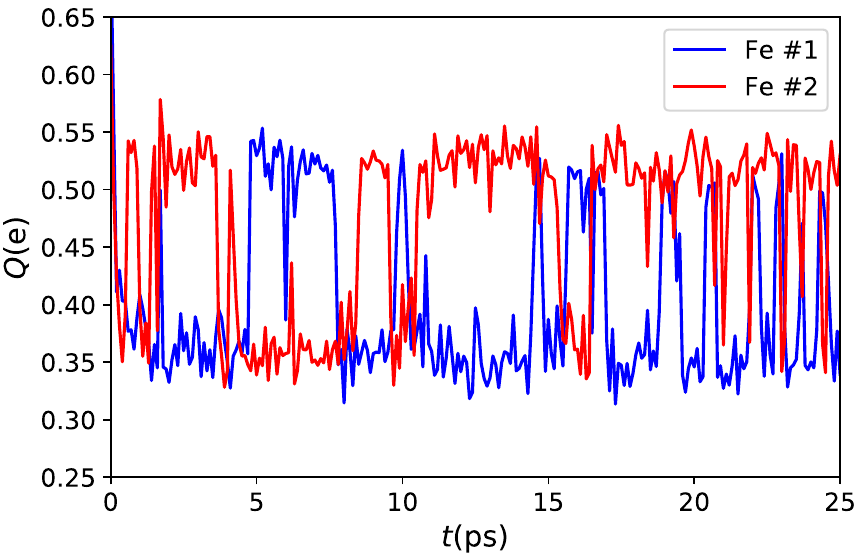}
    \caption{Partial charges $Q$ of the two iron atoms in an MD simulation of the Fe$_2$Cl$_5$ system using the 4G-HDNNP trained to the combined dataset. } 
    \label{fig:redox}
\end{figure}

Here, we restrict ourselves to test if HDNNP 4G(Fe$^{2+}$/Fe$^{3+}$) is in principle able to describe the Fe$_2$Cl$_5$ system in a physically correct way. For this purpose we have performed MD simulations at 300~K of two iron and five chloride ions in a cubic box of water of side length 20~\AA{} containing 258 water molecules. Figure~\ref{fig:redox} shows the predicted Hirshfeld charges of both Fe ions as a function of time, and indeed a splitting into an Fe$^{2+}$ and an Fe$^{3+}$ ion is observed in this trajectory. Moreover, several switches of the oxidation states of both ions can be identified corresponding to an electron transfer from Fe$^{2+}$ to Fe$^{3+}$. These switches in oxidation states between Fe$^{2+}$ and Fe$^{3+}$ and vice versa in many cases occur almost simultaneously such that the overall charge of both iron ions remains approximately constant.
These findings are remarkable and demonstrate that, in spite of the application beyond the range of structures it has been trained to, HDNNP 4G(Fe$^{2+}$/Fe$^{3+}$) is able to describe the Fe$_2$Cl$_5$ system in a qualitatively correct way.

We note that in 4G-HDNNPs the overall charge of the system is conserved by construction by a constraint in the underlying charge equilibration scheme. This charge conservation, however, does not impose any form of constraint on the total charge of the two iron atoms, and in principle, both Fe ions could simultaneously adopt the same Fe$^{3+}$ or Fe$^{2+}$ oxidation state if the excess charge is distributed in other parts of the system. This is indeed observed in several of the trajectories we computed, not only during the rearrangement of solvent molecules upon charge transfer processes but in particular if structural motifs emerge that are far from the training set. This is not a shortcoming of fourth-generation potentials in general, but rather the consequence of a lack of training data in the extrapolation regime in the current parameterization of the potential. Still, interestingly, in all our simulations the charges of the Fe ions remain in the range of Fe$^{2+}$ and Fe$^{3+}$, and basically no charges outside this range have been observed. We conclude that for reliably producing a splitting into an Fe$^{2+}$ and an Fe$^{3+}$ throughout extended simulations, further training data will be needed, which more comprehensively covers the relevant configuration space.

\section{Discussion}\label{sec:discussion}

Using the example of 2G-HDNNPs, the results obtained in this work clearly show that widely-used second-generation MLPs are unable to distinguish chemical species in different oxidation states in environments like liquid water. Moreover, we demonstrate that this limitation can be overcome by 4G-HDNNPs. These findings clearly prove that obtaining both quantitatively and qualitatively accurate results in MLP-based MD simulations hinges on the selection of a suitable machine learning model. This selection requires physical insights into the system, and there is no universal type of MLP that is equally suited for all types of systems providing optimum accuracy and efficiency. While relatively simple 2G-HDNNPs are very efficient and exhibit a close to linear scaling with system size, the charge equilibration step of 4G-HDNNPs increases the computational costs, although more efficient algorithms are becoming available~\citenum{P5932,gubler2024accelerating}. Hence, using MLPs as black box methods and trusting in a high accuracy if just enough training data is provided may lead to qualitatively incorrect simulation outcomes.

An alternative approach to the use of fourth-generation potentials to increase the information about the chemical composition of the system might be the use of modern message passing neural networks~\cite{P5368,P5366,P5577,P6572}. In these potentials, information is passed iteratively from atom to atom, avoiding the use of a fixed cutoff radius of the atomic environments. However, in practice the number of message passing steps is often small to keep the computational costs at a reasonable level. Thus, the information about the system is essentially still local similar to second-generation potentials resulting in the same limitations. Still, message passing might offer an acceptable compromise for systems of moderate size allowing sufficient information transfer.

Another important result of this work is that RMSE values alone cannot be relied on for assessing the quality of an MLP. Large RMSE values of energies and forces are an obvious indication for poor potentials, but low RMSE values do not necessarily guarantee high accuracy. As shown above, the very similar low energy and force RMSEs of the 2G- and 4G-HDNNPs in Table~\ref{tab:RMSE_table} do not allow to predict their qualitatively different performance. Therefore, the common procedure of training new types of MLPs to standard benchmark datasets to assess their quality seems questionable, as lower RMSEs are no evidence for more accurate potentials. Even the long-term stability of MD trajectories is not a sufficient criterion, as they still may provide physically incorrect properties. Therefore, the assessment of the quality of MLPs remains challenging, requiring not only the computation of RMSEs but also the analysis of long trajectories and the determination of a wide range of physical properties.

A possible explanation for the low RMSEs of HDNNP 2G(Fe$^{2+}$/Fe$^{3+}$) might be the different local structure of the iron solvation spheres in the training set. These structural differences could enable the 2G-HDNNP to indirectly infer information about the oxidation states of the Fe ions, underlining the limited value of RMSE values in assessing the quality of MLPs as also pointed out in previous work~\cite{P6346,P6548}. In MD simulations involving thermal fluctuations of the solvation spheres, however, this distinction mechanism breaks down and results in a strongly reduced reliability of 2G-HDNNP energy and force predictions. Moreover, the water molecules in the immediate environment of the iron ions, which interact differently with Fe$^{2+}$ and Fe$^{3+}$, represent only a small fraction of all water molecules in the systems, resulting in a low impact on averaged quantities like RMSEs.

In summary, we conclude that in contrast to local 2G-HDNNPs, non-local 4G-HDNNPs are able to correctly distinguish different oxidation states of ions in solution based on the global chemical composition of the electrolyte, i.e., the number of counter ions present in the system. They provide both the correct solvation structures and ion charges in agreement with DFT. While in the present work a rather simple generalized gradient approximation functional has been employed to compute the reference data, our results are independent of the chosen level of theory and more accurate electronic structure methods can be used following the same procedures in future work. 

The HDNNP training did not include systems with more than one Fe atom and three Cl atoms, and thus lacks information about redox reactions in which an electron is transferred from Fe$^{2+}$ to Fe$^{3+}$. Nevertheless, the present potential appears to be robust enough to describe electron transfer reactions in solution in many cases. Other situations, in which both iron ions temporarily adopt the same oxidation state can be explained by atomic configurations far from the underlying training set. Poor predictions in these situations are expected due to the well-known limited extrapolation capabilities of MLPs, but this issue could be overcome by extending the reference dataset accordingly. 

\section{Methods}\label{sec:methods}

\subsection{Density functional theory calculations}\label{sec:DFT}

The DFT calculations of the small and medium-sized periodic boxes were carried out using the FHI-aims all-electron code~\cite{P2189} employing numerical atomic orbitals as basis functions. Light basis sets and integration grids have been used in the AIMD simulations as well as in the single point calculations in active learning to extend the dataset size.
The Perdew-Burke-Ernzerhof (PBE) generalized gradient approximation functional~\cite{P0010} was employed to describe electronic exchange and correlation in the system, which provides a reasonable description and a clear distinction of Fe$^{2+}$ and Fe$^{3+}$ ions in water that is fully sufficient for the purpose of this work. Open-shell unrestricted calculations have been employed to account for the spin polarization of the iron ions. Relativistic effects were included at the atomic zeroth-order regular approximation (atomic ZORA) \cite{P3082} to accurately describe the heavy Fe atoms. Gaussian smearing with a width of 0.01 eV was used to populate the electronic states. $\Gamma$-point sampling has been found suitable for the medium-sized box. For the small box, a 2$\times$2$\times$2 k-grid has been used to obtain tightly converged total energies that are consistent with the medium-sized systems. The total charge of all structures has been set to zero such that the oxidation states of the iron ions are defined by the number of chlorine atoms. Hirshfeld charges~\cite{P0416} have been computed for all structures for identifying the oxidation states of the Fe ions and for generating the reference charges for training the 4G-HDNNP. Orbital occupations have been checked to ensure that all calculations have converged to the correct electronic states, and the Fe ions have been found in the high-spin state in all cases.

For the initial reference set, short \emph{ab initio} MD trajectories have been run at 300 K in the $NVT$ ensemble employing a Nos\'e-Hoover thermostat~\cite{P2758} and a time step of 0.5 fs starting from different structures using different velocity initializations. After constructing first preliminary potentials active learning has been carried out~\cite{P6548} to select structures with high force uncertainty to expand the dataset until the final potentials have been obtained. The final DFT reference dataset contains in total 29,021 structures, further details about the dataset are given in the SI.

\subsection{High-dimensional neural network potentials}\label{sec:HDNNP}

To date, several generations of high-dimensional neural network potentials (HDNNP) have been introduced, which are applicable to systems with different types of interactions~\cite{P6018}.
In second-generation (2G) HDNNPs~\cite{P1174}, the total energy $E$ of the system is constructed as a sum of atomic energies $E_i$,
\begin{eqnarray}
    E^{\mathrm{2G}}=\sum_{i=1}^{N_{\mathrm{atom}}}E_i \quad , \label{eq:e2g}
\end{eqnarray}
which depend on the local atomic environment up to a cutoff radius $R_{\mathrm{c}}$. The positions of all neighboring atoms inside the resulting cutoff spheres are described by feature vectors of atom-centered symmetry functions (ACSF) as structural descriptors~\cite{P2882}, which are invariant with respect to rotation, translation and permutation to ensure that the potential energy surface inherits these essential properties. For each atom in the system, the respective feature vector is used as input for an atomic feed-forward neural network (NN) providing the atomic energy. To ensure permutation invariance of the total energy, the NN weight parameters and architectures are constrained to be the same for a given chemical element. Moreoever, due to the structure of HDNNPs, the resulting force vectors are equivariant with respect to the relevant symmetry operations of the system. 2G-HDNNPs are local and do not make use of structural information beyond the respective atomic environment. Consequently, their accuracy depends on the size of the cutoff radius, 6~\AA{} in the present work, as well as on the description of the local atomic configurations by the ACSFs.
Apart from 2G-HDNNPS, many other second-generation MLPs have been proposed~\cite{P2630,P4945,P4862,P5794}, which employ different descriptors and machine learning techniques, but are equally based on Eq.~\ref{eq:e2g} and employ a similar cutoff radius for the atomic interactions. Therefore, the results obtained in this work are equally valid for these potentials.

Third-generation HDNNPs~\cite{P2962} include a second set of atomic neural networks predicting local atomic charges as a function of the environment, which are used to additionally compute long-range electrostatic interactions beyond the cutoff. Still, such 3G-HDNNPs are not suited to describe redox reactions in solution, since like 2G-HDNNPs they lack information about counter ions beyond the cutoff radius and thus cannot correctly determine the iron charges and oxidation states.

For this reason, in the present work we have investigated the capabilities of fourth-generation HDNNPs~\cite{P5932} to describe redox processes in solution. Like in 3G-HDNNPs the total energy of 4G-HDNNPs consists of the long-range electrostatic energy and a sum of short-range atomic energies, 
\begin{eqnarray}
    E^{\mathrm{4G}}=E_{\mathrm{short}}+E_{\mathrm{elec}} \quad , \label{eq:4G} 
\end{eqnarray}
but in fourth-generation potentials the atomic partial charges depend on the global structure of the system. Thus, information about the total number of counter ions is available and considered in the determination of the charge distribution in the system. In case of 4G-HDNNPs this is achieved by a global charge equilibration step~\cite{P1448,P4419} based on atomic electronegativities expressed by atomic neural networks. Thus, 4G-HDNNPs are able to describe long-range charge transfer, such as electron transfer from iron atoms to chlorine atom irrespective of their distance in the system, ensuring the formation of ions with the correct oxidation states. 

Such long-range charge transfer is not only important for the calculation of the long-range electrostatic energy but also modifies the charge density and thus the local interactions between all atoms. To take these changes in local bonding into account, in addition to the atom-centered symmetry functions the atomic partial charges are used as additional input descriptors for the atomic energy neural networks. Therefore, the atomic energy contributions can adapt to charge transfer in the system to yield consistent long-range electrostatics and atomic energies, which then are added to the total potential energy (Eq.~\ref{eq:4G}). Further details about the methodology of 2G- and 4G-HDNNPs can be found in Refs.~\citenum{P5932,P6018}.

All HDNNPs have been trained using the RuNNer code~\cite{P5128,P4444} employing the Kalman filter to determine the NN weight parameters~\cite{P1308}. Energies and forces have been used for the training of the atomic energy neural networks, while the weights of the electronegativity NNs determining the charges in the 4G-HDNNP have been optimized to reproduce reference DFT Hirshfeld charges~\cite{P0416}. In case of 4G-HDNNPs the training is a two-step process in which first the atomic charges are learned. In a second step, the electrostatic energies and forces are removed from the respective DFT reference values and the remaining parts of the energies and forces are learned by the atomic energy NNs to avoid double counting of electrostatic energy contributions.

Three datasets containing only the FeCl$_2$ systems, only FeCl$_3$ systems and the combined dataset have been used and randomly split into training (about 90~\%) and test sets (about 10~\%). The compositions of these datasets are given in the SI. The atomic environments are described by ACSFs whose functional forms and parameters are given in the SI. 
The atomic neural networks of all elements contain 2 hidden layers consisting of 20 neurons each. For the neurons in the hidden layers hyperbolic tangent activation functions have been used and for the output neurons linear functions have been employed. The same network architectures and ACSFs have been used for the atomic energy NNs and electronegativity NNs.
L2 regularization \cite{ng2004feature} has been used with the regularization parameter $\lambda$  equal to $10^{-6}$.

\subsection{Molecular dynamics simulations}\label{sec:MD}

Classical MD simulations based on energies and forces provided by the HDNNPs have been run with the LAMMPS code~\cite{P4473} using the n2p2 library for HDNNPs~\cite{P5603}. Simulations in the $NVT$ ensemble were run with a time step of 0.5~fs 
at 300~K employing the Nos\'e-Hoover thermostat \cite{P2758}.

\section{Acknowledgements}

We are grateful for support by the Deutsche Forschungsgemeinschaft (DFG) (BE3264/16-1, project number 495842446 in priority program SPP 2363 ``Utilization and Development of Machine Learning for Molecular Applications – Molecular Machine Learning'') and under Germany's Excellence Strategy – EXC 2033 RESOLV (project-ID 390677874). This research was funded in part by the Austrian Science Fund (FWF) 10.55776/F81. For Open Access purposes, the authors have applied a CC BY public copyright license to any author accepted manuscript version arising from this submission.


%

\end{document}


\setstretch{1.0}

\title{Supporting Information: Machine learning potentials for redox chemistry in solution}

\author{Emir Kocer}
\affiliation{Lehrstuhl f\"ur Theoretische Chemie II, Ruhr-Universit\"at Bochum, 44780 Bochum, Germany} 
\affiliation{Research Center Chemical Sciences and Sustainability, Research Alliance Ruhr, 44780 Bochum, Germany}
\author{Redouan El Haouari}
\affiliation{Lehrstuhl f\"ur Theoretische Chemie II, Ruhr-Universit\"at Bochum, 44780 Bochum, Germany} 
\affiliation{Research Center Chemical Sciences and Sustainability, Research Alliance Ruhr, 44780 Bochum, Germany}
\author{Christoph Dellago}
\affiliation{University of Vienna, Faculty of Physics, Boltzmanngasse 5, A-1090 Vienna, Austria}
\author{J\"{o}rg Behler}
\email{joerg.behler@ruhr-uni-bochum.de}
\affiliation{Lehrstuhl f\"ur Theoretische Chemie II, Ruhr-Universit\"at Bochum, 44780 Bochum, Germany} 
\affiliation{Research Center Chemical Sciences and Sustainability, Research Alliance Ruhr, 44780 Bochum, Germany}

\begin{abstract}

\end{abstract}

\maketitle

\subsection{Reference dataset}

The compositions of the training and test datasets for all six HDNNPs are listed in Table~\ref{tab:dataset}. The combined datasets for 2G(Fe$^{2+}$/Fe$^{3+}$) and 4G(Fe$^{2+}$/Fe$^{3+}$) contain a few additional structures from active learning compared to the individual subsets.

\begin{table}[h] 
\centering
\caption{Numbers of structures in the training and test datasets of all six HDNNPs constructed in this work. }
\label{tab:dataset}
{\footnotesize 
\begin{tabular}{@{}l|c|c|c|c|c|c|c@{}} 
\toprule
& \multicolumn{2}{c|}{2G(Fe$^{2+}$)} & \multicolumn{2}{c|}{4G(Fe$^{2+}$)}\\
 & Training & Test & Training & Test \\
\hline
small box  & 5753 & 665 & 5753 & 665 \\
medium box & 7603 & 803 & 7603 & 803\\
total  & 13356 & 1468 & 13356 & 1468 \\
\hline
\multicolumn{5}{c}{}\\

& \multicolumn{2}{c|}{2G(Fe$^{3+}$)} & \multicolumn{2}{c|}{4G(Fe$^{3+}$)}\\
 & Training & Test & Training & Test \\
\hline
small box & 7433  & 843 & 7433 & 843\\
medium box & 4907 & 518 & 4907 & 518 \\
total  & 12340 & 1361 & 12340 &1361 \\
\hline
\multicolumn{5}{c}{}\\

& \multicolumn{2}{c|}{2G(Fe$^{2+}$/Fe$^{3+}$)} & \multicolumn{2}{c|}{4G(Fe$^{2+}$/Fe$^{3+}$)}\\
 & Training & Test & Training & Test \\
\hline
small box & 13170 & 1484 & 13170 & 1484\\
medium box & 12859 & 1468 & 12859 & 1468 \\
total  & 26069 & 2952 & 26069 & 2952\\
\hline
\bottomrule
\end{tabular}
} 
\end{table}

\subsection{HDNNP training}

All HDNNPs in this study have been trained with the RuNNer code~\cite{P5128,P4444}. Local atomic environments are encoded using radial and angular atom-centered symmetry functions (ACSF) \cite{behler2011atom}. \\
Radial ACSFs can be considered as continuous coordination numbers and are used to encode the radial distribution of neighbors of an atom as a function of their distance up to the cutoff radius. They are calculated via 
\begin{eqnarray}
    G_{\mathrm{radial},i}=\sum_{j\in R_{\mathrm{c}}} \mathrm{e}^{-\eta(R_{ij}-R_{s})^{2}}\cdot f_{c}(R_{ij}) \quad , \label{eq:radial}
\end{eqnarray}
where $\eta$ and $R_s$ are hyperparameters defining the width and radial displacement of a Gaussian sphere around the central atom. Angular ACSFs are used to encode the angular distribution of neighbors. They are calculated via  \ref{eq:angular}
\begin{equation}
\begin{aligned}
    G_{\mathrm{angular},i}=2^{1-\zeta}\sum_{j\in R_{\mathrm{c}}}\sum_{k\neq j\in R_{\mathrm{c}}}(1+\lambda \mathrm{cos}\theta_{ijk})^{\zeta}\cdot \\
    \quad  e^{-\eta(R_{ij}^{2}+R_{ik}^{2}+R_{jk}^{2})} \cdot f_{c}(R_{ij})f_{c}(R_{ik})f_{c}(R_{jk}) \label{eq:angular}
\end{aligned}
\end{equation}
where $\theta_{ijk}$ is the angle enclosed by the atomic connections $ij$ and $ik$ of central atom $i$, and $\zeta$, $\lambda$ and $\eta$ are hyperparameters.

The cutoff function $f_{\mathrm{c}}(R_{ij})$ is defined as 
\begin{equation}
\begin{aligned}
    f_{\mathrm{c}}(R_{ij}) = \text{tanh}^3\left( 1 - \frac{R_{ij}}{R_{\mathrm{c}}} \right)  \quad , \label{eq:cutoff}
\end{aligned}
\end{equation}
where $R_{\mathrm{c}}$ is the cutoff radius that has been selected as 6~$\text{\AA}$ for all symmetry functions. 
\\
The radial ACSFs of all central elements that have been used in this study are listed in Tables \ref{tab:radial_symfunction_hydrogen}, \ref{tab:radial_symfunction_oxygen}, \ref{tab:radial_symfunction_chlorine} and \ref{tab:radial_symfunction_iron}, whereas the angular ACSFs are given in Tables \ref{tab:angular_symfunction_hydrogen}, \ref{tab:angular_symfunction_oxygen}, \ref{tab:angular_symfunction_chlorine} and \ref{tab:angular_symfunction_iron}.

\begin{table}[t]
\centering
\caption{Parameters of the radial symmetry functions of hydrogen, where $\eta$ (1/Bohr$^2$) and $R_{\mathbf{s}}$ (Bohr) are defined in Eq.\ \ref{eq:radial}.}\label{tab:radial_symfunction_hydrogen}
{\footnotesize
\begin{tabular}{c|cccccc}
        \toprule
        ID & Atom & Neighbor & $\eta$ & $R_{\mathbf{s}}$  \\
        \hline
        \midrule
        1  & H  & H  & 0.0010 & 0.0 \\
        2  & H  & H  & 0.0100 & 0.0 \\
        3 & H  & H  & 0.0300 & 0.0 \\
        4 & H  & H  & 0.0600 & 0.0 \\
        5 & H  & H  & 0.1500 & 1.9 \\
        6 & H  & H  & 0.3000 & 1.9 \\
        7 & H  & H  & 0.6000 & 1.9 \\
        8 & H  & H  & 1.5000 & 1.9 \\
        9  & H  & O  & 0.0010 & 0.0 \\
        10  & H  & O  & 0.0100 & 0.0 \\
        11 & H  & O  & 0.0300 & 0.0 \\
        12 & H  & O  & 0.0600 & 0.0 \\
        13 & H  & O  & 0.1500 & 0.9 \\
        14 & H  & O  & 0.3000 & 0.9 \\
        15 & H  & O  & 0.6000 & 0.9 \\
        16 & H  & O  & 1.5000 & 0.9 \\
        17  & H  & Cl & 0.0005 & 0.0 \\
        18  & H  & Cl & 0.0085 & 0.0 \\
        19 & H  & Cl & 0.0250 & 0.0 \\
        20 & H  & Cl & 0.0500 & 0.0 \\
        21 & H  & Cl & 0.1000 & 0.9 \\
        22 & H  & Cl & 0.5000 & 0.9 \\
        23 & H  & Cl & 1.2500 & 0.9 \\
        24  & H  & Fe & 0.0002 & 0.0 \\
        25  & H  & Fe & 0.0065 & 0.0 \\
        26  & H  & Fe & 0.0150 & 0.0 \\
        27 & H  & Fe & 0.0400 & 0.0 \\
        28 & H  & Fe & 0.0750 & 0.9 \\
        29 & H  & Fe & 0.4000 & 0.9 \\
        30 & H  & Fe & 1.2000 & 0.9 \\
        \bottomrule
    \end{tabular}
}
\end{table}

\begin{table}[t]
\centering
\caption{Parameters of the radial symmetry functions of oxygen, where $\eta$ (1/Bohr$^2$) and $R_{\mathbf{s}}$ (Bohr) are defined in Eq.\ \ref{eq:radial}.}
{\footnotesize
\begin{tabular}{c|cccccc}
        \toprule
        ID & Atom & Neighbor & $\eta$ & $R_{\mathbf{s}}$  \\
        \hline
        \midrule
        1  & O  & H  & 0.0010 & 0.0 \\
        2  & O  & H  & 0.0100 & 0.0 \\
        3 & O  & H  & 0.0300 & 0.0 \\
        4 & O  & H  & 0.0600 & 0.0 \\
        5 & O  & H  & 0.1500 & 0.9 \\
        6 & O  & H  & 0.3000 & 0.9 \\
        7 & O  & H  & 0.6000 & 0.9 \\
        8 & O  & H  & 1.5000 & 0.9 \\
        
        9  & O  & O  & 0.0010 & 0.0 \\
        10  & O  & O  & 0.0100 & 0.0 \\
        11 & O  & O  & 0.0300 & 0.0 \\
        12 & O  & O  & 0.0600 & 0.0 \\
        13 & O  & O  & 0.1500 & 0.9 \\
        14 & O  & O  & 0.3000 & 0.9 \\
        15 & O  & O  & 0.6000 & 0.9 \\
        16 & O  & O  & 1.5000 & 0.9 \\
        
        17  & O  & Cl & 0.0075 & 0.0 \\
        18  & O  & Cl & 0.0200 & 0.0 \\
        19 & O  & Cl & 0.0300 & 0.0 \\
        20 & O  & Cl & 0.0500 & 0.0 \\
        21 & O  & Cl & 0.2500 & 0.9 \\
        
        22  & O  & Fe & 0.0002 & 0.0 \\
        23  & O  & Fe & 0.0065 & 0.0 \\
        24  & O  & Fe & 0.0150 & 0.0 \\
        25 & O  & Fe & 0.0400 & 0.0 \\
        26 & O  & Fe & 0.0750 & 0.9 \\
        27 & O  & Fe & 0.4000 & 0.9 \\
        28 & O  & Fe & 1.2000 & 0.9 \\
        \bottomrule
    \end{tabular}
}
\label{tab:radial_symfunction_oxygen}
\end{table}

\begin{table}[t]
\centering
\caption{Parameters of the radial symmetry functions of chlorine, where $\eta$ (1/Bohr$^2$) and $R_{\mathbf{s}}$ (Bohr) are defined in Eq.\ \ref{eq:radial}.}
{\footnotesize
\begin{tabular}{c|cccccc}
        \toprule
        ID & Atom & Neighbor & $\eta$ & $R_{\mathbf{s}}$  \\
        \hline
        \midrule
        1  & Cl  & H  & 0.0005 & 0.0 \\
        2  & Cl  & H  & 0.0085 & 0.0 \\
        3 & Cl  & H  & 0.0250 & 0.0 \\
        4 & Cl  & H  & 0.0500 & 0.0 \\
        5 & Cl  & H  & 0.1000 & 0.9 \\
        6 & Cl  & H  & 0.5000 & 0.9 \\
        7 & Cl  & H  & 1.2500 & 0.9 \\
        
        8  & Cl  & O  & 0.0075 & 0.0 \\
        9  & Cl  & O  & 0.0200 & 0.0 \\
        10 & Cl  & O  & 0.0300 & 0.0 \\
        11 & Cl  & O  & 0.0500 & 0.0 \\
        12 & Cl  & O  & 0.2500 & 0.9 \\
        13  & Cl  & Cl & 0.0001 & 0.0 \\
        14  & Cl  & Cl & 0.0010 & 0.0 \\
        15 & Cl  & Cl & 0.0050 & 0.0 \\
        16 & Cl  & Cl & 0.0100 & 0.9 \\
        17 & Cl  & Cl & 0.2500 & 0.9 \\
        18  & Cl  & Fe & 0.0001 & 0.0 \\
        19  & Cl  & Fe & 0.0008 & 0.0 \\
        20  & Cl  & Fe & 0.0040 & 0.0 \\
        21 & Cl  & Fe & 0.0100 & 0.9 \\
       
        \bottomrule
    \end{tabular}
}
\label{tab:radial_symfunction_chlorine}
\end{table}

\begin{table}[t]
\centering
\caption{Parameters of the radial symmetry functions of iron, where $\eta$ (1/Bohr$^2$) and $R_{\mathbf{s}}$ (Bohr) are defined in Eq.\ \ref{eq:radial}.}
{\footnotesize
\begin{tabular}{c|cccccc}
        \toprule
        ID & Atom & Neighbor & $\eta$ & $R_{\mathbf{s}}$  \\
        \hline
        \midrule
        1  & Fe  & H  & 0.0002 & 0.0 \\
        2  & Fe  & H  & 0.0065 & 0.0 \\
        3 & Fe  & H  & 0.0150 & 0.0 \\
        4 & Fe  & H  & 0.0400 & 0.0 \\
        5 & Fe  & H  & 0.0750 & 0.9 \\
        6 & Fe  & H  & 0.4000 & 0.9 \\
        7 & Fe  & H  & 1.2000 & 0.9 \\
        8  & Fe  & O  & 0.0002 & 0.0 \\
        9  & Fe  & O  & 0.0065 & 0.0 \\
        10 & Fe  & O  & 0.0150 & 0.0 \\
        11 & Fe  & O  & 0.0400 & 0.0 \\
        12 & Fe  & O  & 0.0750 & 0.9 \\
        13  & Fe  & O & 0.4000 & 0.9 \\
        14  & Fe  & O & 1.2000 & 0.9 \\
        15 & Fe  & Cl & 0.0001 & 0.0 \\
        16 & Fe  & Cl & 0.0008 & 0.0 \\
        17 & Fe  & Cl & 0.0040 & 0.0 \\
        18 & Fe  & Cl & 0.0100 & 0.9 \\
        \bottomrule
    \end{tabular}
}
\label{tab:radial_symfunction_iron}
\end{table}

\begin{table}[t]
\centering
\caption{Parameter of the angular ACSFs of hydrogen, where $\eta$ (1/Bohr$^2$), $\lambda$ and $\zeta$ are defined in Eq.\ \ref{eq:angular}.}
{\footnotesize
\begin{tabular}{c|cccccccc}
    \toprule
    ID & Atom & Neighbor 1 & Neighbor 2 & $\eta$ & $\lambda$ & $\zeta$\\
    \hline
    \midrule
    1  & H  & H  & H  & 0.00000  & -1.00000  & 1.00000   \\
    2  & H  & H  & H  & 0.00000  &  1.00000  & 1.00000   \\
    3  & H  & H  & H  & 0.00000  & -1.00000  & 4.00000   \\
    4  & H  & H  & H  & 0.00000  &  1.00000  & 4.00000   \\
    5  & H  & H  & H  & 0.04496  & -1.00000  & 1.00000   \\
    6  & H  & H  & H  & 0.04496  &  1.00000  & 1.00000   \\
    7  & H  & H  & H  & 0.04496  & -1.00000  & 4.00000   \\
    8  & H  & H  & H  & 0.04496  &  1.00000  & 4.00000   \\  
    
    9  & H  & H  & O  & 0.00000  & -1.00000  & 1.00000   \\
    10 & H  & H  & O  & 0.00000  &  1.00000  & 1.00000   \\
    11 & H  & H  & O  & 0.00000  & -1.00000  & 4.00000   \\
    12 & H  & H  & O  & 0.00000  &  1.00000  & 4.00000   \\
    13 & H  & H  & O  & 0.05370  & -1.00000  & 1.00000   \\
    14 & H  & H  & O  & 0.05370  &  1.00000  & 1.00000   \\
    15 & H  & H  & O  & 0.05370  & -1.00000  & 4.00000   \\
    16 & H  & H  & O  & 0.05370  &  1.00000  & 4.00000   \\  
    
    17 & H  & O  & O  & 0.00000  & -1.00000  & 1.00000   \\
    18 & H  & O  & O  & 0.00000  &  1.00000  & 1.00000   \\
    19 & H  & O  & O  & 0.00000  & -1.00000  & 4.00000   \\
    20 & H  & O  & O  & 0.00000  &  1.00000  & 4.00000   \\
    21 & H  & O  & O  & 0.04254  & -1.00000  & 1.00000   \\
    22 & H  & O  & O  & 0.04254  &  1.00000  & 1.00000   \\
    23 & H  & O  & O  & 0.04254  & -1.00000  & 4.00000   \\
    24 & H  & O  & O  & 0.04254  &  1.00000  & 4.00000   \\
    
    25 & H  & H  & Cl  & 0.00000  & -1.00000  & 1.00000   \\
    26 & H  & H  & Cl  & 0.00000  &  1.00000  & 1.00000   \\
    27 & H  & H  & Cl  & 0.00000  & -1.00000  & 4.00000   \\
    28 & H  & H  & Cl  & 0.00000  &  1.00000  & 4.00000   \\
    29 & H  & H  & Cl & 0.02919  & -1.00000  & 1.00000   \\
    30 & H  & H  & Cl & 0.02919  &  1.00000  & 1.00000   \\
    31 & H  & H  & Cl & 0.02919  & -1.00000  & 4.00000   \\
    32 & H  & H  & Cl & 0.02919  &  1.00000  & 4.00000   \\ 
    
    33 & H  & O  & Cl  & 0.00000  & -1.00000  & 1.00000   \\
    34 & H  & O  & Cl  & 0.00000  &  1.00000  & 1.00000   \\
    35 & H  & O  & Cl  & 0.00000  & -1.00000  & 4.00000   \\
    36 & H  & O  & Cl  & 0.00000  &  1.00000  & 4.00000   \\
    37 & H  & O  & Cl & 0.02919  & -1.00000  & 1.00000   \\
    38 & H  & O  & Cl & 0.02919  &  1.00000  & 1.00000   \\
    39 & H  & O  & Cl & 0.02919  & -1.00000  & 4.00000   \\
    40 & H  & O  & Cl & 0.02919  &  1.00000  & 4.00000   \\ 
    
    41 & H  & Cl & Cl & 0.00000  & -1.00000  & 1.00000   \\
    42 & H  & Cl & Cl & 0.00000  &  1.00000  & 1.00000   \\
    43 & H  & Cl & Cl & 0.00000  & -1.00000  & 4.00000   \\
    44 & H  & Cl & Cl & 0.00000  &  1.00000  & 4.00000   \\
    45 & H  & Cl & Cl & 0.01707  &  1.00000  & 1.00000   \\
    46 & H  & Cl & Cl & 0.01707  &  1.00000  & 4.00000   \\
    
    47 & H  & H  & Fe  & 0.00000  & -1.00000  & 1.00000   \\
    48 & H  & H  & Fe  & 0.00000  &  1.00000  & 1.00000   \\
    49 & H  & H  & Fe  & 0.00000  & -1.00000  & 4.00000   \\
    50 & H  & H  & Fe  & 0.00000  &  1.00000  & 4.00000   \\
    51 & H  & H  & Fe & 0.02973  & -1.00000  & 1.00000   \\
    52 & H  & H  & Fe & 0.02973  &  1.00000  & 1.00000   \\
    53 & H  & H  & Fe & 0.02973  & -1.00000  & 4.00000   \\
    54 & H  & H  & Fe & 0.02973  &  1.00000  & 4.00000   \\
    
    55 & H  & O  & Fe & 0.00000  & -1.00000  & 1.00000   \\
    56 & H  & O  & Fe & 0.00000  &  1.00000  & 1.00000   \\
    57 & H  & O  & Fe & 0.00000  & -1.00000  & 4.00000   \\
    58 & H  & O  & Fe & 0.00000  &  1.00000  & 4.00000   \\
    59 & H  & O  & Fe & 0.02973  & -1.00000  & 1.00000   \\
    60 & H  & O  & Fe & 0.02973  &  1.00000  & 1.00000   \\
    61 & H  & O  & Fe & 0.02973  & -1.00000  & 4.00000   \\
    62 & H  & O  & Fe & 0.02973  &  1.00000  & 4.00000   \\
    
    63 & H  & Cl & Fe & 0.00000  & -1.00000  & 1.00000   \\
    64 & H  & Cl & Fe & 0.00000  &  1.00000  & 1.00000   \\
    65 & H  & Cl & Fe & 0.00000  & -1.00000  & 4.00000   \\
    66 & H  & Cl & Fe & 0.00000  &  1.00000  & 4.00000   \\
   
    \bottomrule
\end{tabular}
}
\label{tab:angular_symfunction_hydrogen}
\end{table}

\begin{table}[t]
\centering
\caption{Parameter of the angular ACSFs of oxygen, where $\eta$ (1/Bohr$^2$), $\lambda$ and $\zeta$ are defined in Eq.\ \ref{eq:angular}.}
{\footnotesize
\begin{tabular}{c|cccccccc}
    \toprule
    ID & Atom & Neighbor 1 & Neighbor 2 & $\eta$ & $\lambda$ & $\zeta$\\
    \hline
    \midrule
    1  & O  & H  & H  & 0.00000  & -1.00000  & 1.00000   \\
    2  & O  & H  & H  & 0.00000  &  1.00000  & 1.00000   \\
    3  & O  & H  & H  & 0.00000  & -1.00000  & 4.00000   \\
    4  & O  & H  & H  & 0.00000  &  1.00000  & 4.00000   \\
    5  & O  & H  & H  & 0.05370  & -1.00000  & 1.00000   \\
    6  & O  & H  & H  & 0.05370  &  1.00000  & 1.00000   \\
    7  & O  & H  & H  & 0.05370  & -1.00000  & 4.00000   \\
    8  & O  & H  & H  & 0.05370  &  1.00000  & 4.00000   \\  
    
    9  & O  & H  & O  & 0.00000  & -1.00000  & 1.00000   \\
    10 & O  & H  & O  & 0.00000  &  1.00000  & 1.00000   \\
    11 & O  & H  & O  & 0.00000  & -1.00000  & 4.00000   \\
    12 & O  & H  & O  & 0.00000  &  1.00000  & 4.00000   \\
    13 & O  & H  & O  & 0.04254  & -1.00000  & 1.00000   \\
    14 & O  & H  & O  & 0.04254  &  1.00000  & 1.00000   \\
    15 & O  & H  & O  & 0.04254  & -1.00000  & 4.00000   \\
    16 & O  & H  & O  & 0.04254  &  1.00000  & 4.00000   \\  
    
    17 & O  & O  & O  & 0.00000  & -1.00000  & 1.00000   \\
    18 & O  & O  & O  & 0.00000  &  1.00000  & 1.00000   \\
    19 & O  & O  & O  & 0.00000  & -1.00000  & 4.00000   \\
    20 & O  & O  & O  & 0.00000  &  1.00000  & 4.00000   \\
    21 & O  & O  & O  & 0.01145  & -1.00000  & 1.00000   \\
    22 & O  & O  & O  & 0.01145  &  1.00000  & 1.00000   \\
    23 & O  & O  & O  & 0.01145  & -1.00000  & 4.00000   \\
    24 & O  & O  & O  & 0.01145  &  1.00000  & 4.00000   \\
    
    25 & O  & H  & Cl  & 0.00000  & -1.00000  & 1.00000   \\
    26 & O  & H  & Cl  & 0.00000  &  1.00000  & 1.00000   \\
    27 & O  & H  & Cl  & 0.00000  & -1.00000  & 4.00000   \\
    28 & O  & H  & Cl  & 0.00000  &  1.00000  & 4.00000   \\
    29 & O  & H  & Cl & 0.02973  & -1.00000  & 1.00000   \\
    30 & O  & H  & Cl & 0.02973  &  1.00000  & 1.00000   \\
    31 & O  & H  & Cl & 0.02973  & -1.00000  & 4.00000   \\
    32 & O  & H  & Cl & 0.02973  &  1.00000  & 4.00000   \\ 
    
    33 & O  & O  & Cl  & 0.00000  & -1.00000  & 1.00000   \\
    34 & O  & O  & Cl  & 0.00000  &  1.00000  & 1.00000   \\
    35 & O  & O  & Cl  & 0.00000  & -1.00000  & 4.00000   \\
    36 & O  & O  & Cl  & 0.00000  &  1.00000  & 4.00000   \\
    37 & O  & O  & Cl & 0.01035  & -1.00000  & 1.00000   \\
    38 & O  & O  & Cl & 0.01035  &  1.00000  & 1.00000   \\
    39 & O  & O  & Cl & 0.01035  & -1.00000  & 4.00000   \\
    40 & O  & O  & Cl & 0.01035  &  1.00000  & 4.00000   \\ 
    
    41 & O  & Cl & Cl & 0.00941  & -1.00000  & 1.00000   \\
    42 & O  & Cl & Cl & 0.00000  &  1.00000  & 1.00000   \\
    43 & O  & Cl & Cl & 0.00000  & -1.00000  & 4.00000   \\
    44 & O  & Cl & Cl & 0.00000  &  1.00000  & 4.00000   \\
    45 & O  & Cl & Cl & 0.00941  &  1.00000  & 1.00000   \\
    46 & O  & Cl & Cl & 0.00941  &  1.00000  & 4.00000   \\
    
    47 & O  & H  & Fe  & 0.00000  & -1.00000  & 1.00000   \\
    48 & O  & H  & Fe  & 0.00000  &  1.00000  & 1.00000   \\
    49 & O  & H  & Fe  & 0.00000  & -1.00000  & 4.00000   \\
    50 & O  & H  & Fe  & 0.00000  &  1.00000  & 4.00000   \\
    51 & O  & H  & Fe & 0.02973  & -1.00000  & 1.00000   \\
    52 & O  & H  & Fe & 0.02973  &  1.00000  & 1.00000   \\
    53 & O  & H  & Fe & 0.02973  & -1.00000  & 4.00000   \\
    54 & O  & H  & Fe & 0.02973  &  1.00000  & 4.00000   \\
    
    55 & O  & O  & Fe & 0.00000  & -1.00000  & 1.00000   \\
    56 & O  & O  & Fe & 0.00000  &  1.00000  & 1.00000   \\
    57 & O  & O  & Fe & 0.00000  & -1.00000  & 4.00000   \\
    58 & O  & O  & Fe & 0.00000  &  1.00000  & 4.00000   \\
    59 & O  & O  & Fe & 0.02973  & -1.00000  & 1.00000   \\
    60 & O  & O  & Fe & 0.02973  &  1.00000  & 1.00000   \\
    61 & O  & O  & Fe & 0.02973  & -1.00000  & 4.00000   \\
    62 & O  & O  & Fe & 0.02973  &  1.00000  & 4.00000   \\
    
    63 & O  & Cl & Fe & 0.00000  & -1.00000  & 1.00000   \\
    64 & O  & Cl & Fe & 0.00000  &  1.00000  & 1.00000   \\
    65 & O  & Cl & Fe & 0.00000  & -1.00000  & 4.00000   \\
   
    \bottomrule
\end{tabular}
}
\label{tab:angular_symfunction_oxygen}
\end{table}

\begin{table}[t]
\centering
\caption{Parameter of the angular ACSFs of chlorine, where $\eta$ (1/Bohr$^2$), $\lambda$ and $\zeta$ are defined in Eq.\ \ref{eq:angular}.}
{\footnotesize
\begin{tabular}{c|cccccccc}
    \toprule
    ID & Atom & Neighbor 1 & Neighbor 2 & $\eta$ & $\lambda$ & $\zeta$\\
    \hline
    \midrule
    1  & Cl  & H  & H  & 0.00000  & -1.00000  & 1.00000   \\
    2  & Cl  & H  & H  & 0.00000  &  1.00000  & 1.00000   \\
    3  & Cl  & H  & H  & 0.00000  & -1.00000  & 4.00000   \\
    4  & Cl  & H  & H  & 0.00000  &  1.00000  & 4.00000   \\
    5  & Cl  & H  & H  & 0.05370  & -1.00000  & 1.00000   \\
    6  & Cl  & H  & H  & 0.05370  &  1.00000  & 1.00000   \\
    7  & Cl  & H  & H  & 0.05370  & -1.00000  & 4.00000   \\
    8  & Cl  & H  & H  & 0.05370  &  1.00000  & 4.00000   \\  
    
    9  & Cl  & H  & O  & 0.00000  & -1.00000  & 1.00000   \\
    10 & Cl  & H  & O  & 0.00000  &  1.00000  & 1.00000   \\
    11 & Cl  & H  & O  & 0.00000  & -1.00000  & 4.00000   \\
    12 & Cl  & H  & O  & 0.00000  &  1.00000  & 4.00000   \\
    13 & Cl  & H  & O  & 0.04254  & -1.00000  & 1.00000   \\
    14 & Cl  & H  & O  & 0.04254  &  1.00000  & 1.00000   \\
    15 & Cl  & H  & O  & 0.04254  &  1.00000  & 4.00000   \\  
    
    16 & Cl  & O  & O  & 0.00000  & -1.00000  & 1.00000   \\
    17 & Cl  & O  & O  & 0.00000  &  1.00000  & 1.00000   \\
    18 & Cl  & O  & O  & 0.00000  & -1.00000  & 4.00000   \\
    19 & Cl  & O  & O  & 0.00000  &  1.00000  & 4.00000   \\
    20 & Cl  & O  & O  & 0.01145  & -1.00000  & 1.00000   \\
    21 & Cl  & O  & O  & 0.01145  &  1.00000  & 1.00000   \\
    22 & Cl  & O  & O  & 0.01145  &  1.00000  & 4.00000   \\
    
    23 & Cl  & H  & Cl  & 0.00000  & -1.00000  & 1.00000   \\
    24 & Cl  & H  & Cl  & 0.00000  &  1.00000  & 1.00000   \\
    25 & Cl  & H  & Cl  & 0.00000  & -1.00000  & 4.00000   \\
    26 & Cl  & H  & Cl  & 0.00000  &  1.00000  & 4.00000   \\
    27 & Cl  & H  & Cl & 0.02973  & -1.00000  & 1.00000   \\
    28 & Cl  & H  & Cl & 0.02973  &  1.00000  & 1.00000   \\
    29 & Cl  & H  & Cl & 0.02973  &  1.00000  & 4.00000   \\ 
    
    30 & Cl  & O  & Cl  & 0.00000  & -1.00000  & 1.00000   \\
    31 & Cl  & O  & Cl  & 0.00000  &  1.00000  & 1.00000   \\
    32 & Cl  & O  & Cl  & 0.00000  &  1.00000  & 4.00000   \\
    33 & Cl  & O  & Cl & 0.01035  & -1.00000  & 1.00000   \\
    34 & Cl  & O  & Cl & 0.01035  &  1.00000  & 1.00000   \\
    35 & Cl  & O  & Cl & 0.01035  &  1.00000  & 4.00000   \\ 
    
    36 & Cl  & H  & Fe  & 0.00000  & -1.00000  & 1.00000   \\
    37 & Cl  & H  & Fe  & 0.00000  & 1.00000  & 1.00000   \\
    38 & Cl  & H  & Fe  & 0.02973  & 1.00000  & 1.00000   \\
    39 & Cl  & H  & Fe  & 0.00000  & 1.00000  & 4.00000   \\
    40 & Cl  & H  & Fe  & 0.02973  & 1.00000  & 4.00000   \\
    
    41 & Cl  & O  & Fe & 0.00000  & -1.00000  & 1.00000   \\
    42 & Cl  & O  & Fe & 0.00000  &  1.00000  & 1.00000   \\
    43 & Cl  & O  & Fe & 0.00000  &  1.00000  & 4.00000   \\

    \bottomrule
\end{tabular}
}
\label{tab:angular_symfunction_chlorine}
\end{table}

\begin{table}[t]
\centering
\caption{Parameter of the angular ACSFs of iron, where $\eta$ (1/Bohr$^2$), $\lambda$ and $\zeta$ are defined in Eq.\ \ref{eq:angular}.}
{\footnotesize
\begin{tabular}{c|cccccccc}
    \toprule
    ID & Atom & Neighbor 1 & Neighbor 2 & $\eta$ & $\lambda$ & $\zeta$\\
    \hline
    \midrule
    1  & Fe  & H  & H  & 0.00000  & -1.00000  & 1.00000   \\
    2  & Fe  & H  & H  & 0.00000  &  1.00000  & 1.00000   \\
    3  & Fe  & H  & H  & 0.00000  & -1.00000  & 4.00000   \\
    4  & Fe  & H  & H  & 0.00000  &  1.00000  & 4.00000   \\
    5  & Fe  & H  & H  & 0.05370  & -1.00000  & 1.00000   \\
    6  & Fe  & H  & H  & 0.05370  &  1.00000  & 1.00000   \\
    7  & Fe  & H  & H  & 0.05370  &  1.00000  & 4.00000   \\  
    
    8  & Fe  & H  & O  & 0.00000  & -1.00000  & 1.00000   \\
    9 & Fe  & H  & O  & 0.00000  &  1.00000  & 1.00000   \\
    10 & Fe  & H  & O  & 0.00000  & -1.00000  & 4.00000   \\
    11 & Fe  & H  & O  & 0.00000  &  1.00000  & 4.00000   \\
    12 & Fe  & H  & O  & 0.04254  & -1.00000  & 1.00000   \\
    13 & Fe  & H  & O  & 0.04254  &  1.00000  & 1.00000   \\
    14 & Fe  & H  & O  & 0.04254  &  -1.00000  & 4.00000   \\
    15 & Fe  & H  & O  & 0.04254  &  1.00000  & 4.00000   \\  
    
    16 & Fe  & O  & O  & 0.00000  & -1.00000  & 1.00000   \\
    17 & Fe  & O  & O  & 0.00000  &  1.00000  & 1.00000   \\
    18 & Fe  & O  & O  & 0.00000  & -1.00000  & 4.00000   \\
    19 & Fe  & O  & O  & 0.00000  &  1.00000  & 4.00000   \\
    20 & Fe  & O  & O  & 0.01145  & -1.00000  & 1.00000   \\
    21 & Fe  & O  & O  & 0.01145  &  1.00000  & 1.00000   \\
    22 & Fe  & O  & O  & 0.01145  & -1.00000  & 4.00000   \\
    23 & Fe  & O  & O  & 0.01145  &  1.00000  & 4.00000   \\
    
    24 & Fe  & H  & Cl  & 0.00000  &  1.00000  & 1.00000   \\
    25 & Fe  & H  & Cl  & 0.00000  &  1.00000  & 4.00000   \\
    
    26 & Fe  & O  & Cl  & 0.00000  & -1.00000  & 1.00000   \\
    27 & Fe  & O  & Cl  & 0.00000  &  1.00000  & 1.00000   \\
    28 & Fe  & O  & Cl  & 0.00000  &  1.00000  & 4.00000   \\
    29 & Fe  & O  & Cl & 0.01035  & -1.00000  & 1.00000   \\
    30 & Fe  & O  & Cl & 0.01035  &  1.00000  & 1.00000   \\
    31 & Fe  & O  & Cl & 0.01035  &  1.00000  & 4.00000   \\

    \bottomrule
\end{tabular}
}
\label{tab:angular_symfunction_iron}
\end{table}

\subsection{Error correlations}
\begin{figure*}[!]
    \centering
    \includegraphics[width=0.4\textwidth]{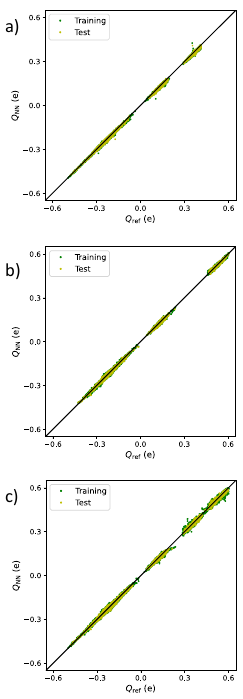}
    \caption{Correlation of the 4G-HDNNP charges $Q_{\mathrm{NN}}$ and the DFT reference charges $Q_{\mathrm{ref}}$ for HDNNP 4G(Fe$^{2+}$) (a), HDNNP 4G(Fe$^{3+}$) (b), and HDNNP 4G(Fe$^{2+}$/Fe$^{3+}$) (c). }
    \label{fig:charge_correlation}
\end{figure*}

\begin{figure*}[!]
    \centering
    \includegraphics[width=0.75\textwidth]{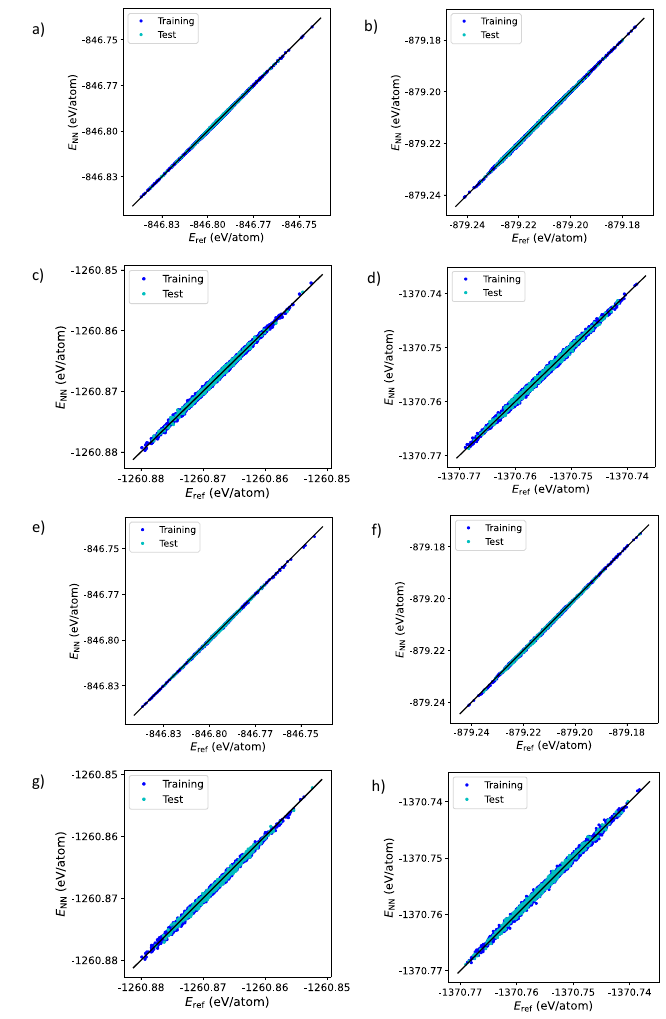}
    \caption{Correlation of the 2G-HDNNP energies $E_{\mathrm{NN}}$ and the DFT energies $E_{\mathrm{ref}}$ per atom for medium-sized FeCl$_2$ structures predicted by 2G(Fe$^{2+}$) (a), medium-sized FeCl$_2$ structures predicted by 2G(Fe$^{2+}$/Fe$^{3+}$) (b), small FeCl$_2$ structures predicted by 2G(Fe$^{2+}$) (c), small FeCl$_2$ structures predicted by 2G(Fe$^{2+}$/Fe$^{3+}$) (d), medium-sized FeCl$_3$ structures predicted by 2G(Fe$^{3+}$) (e), medium-sized FeCl$_3$ structures predicted by 2G(Fe$^{2+}$/Fe$^{3+}$) (f), small FeCl$_3$ structures predicted by 2G(Fe$^{3+}$) (g) and small FeCl$_3$ structures predicted by 2G(Fe$^{2+}$/Fe$^{3+}$) (h). }
    \label{fig:energy_correlation2G}
\end{figure*}

\begin{figure*}[!]
    \centering
    \includegraphics[width=0.75\textwidth]{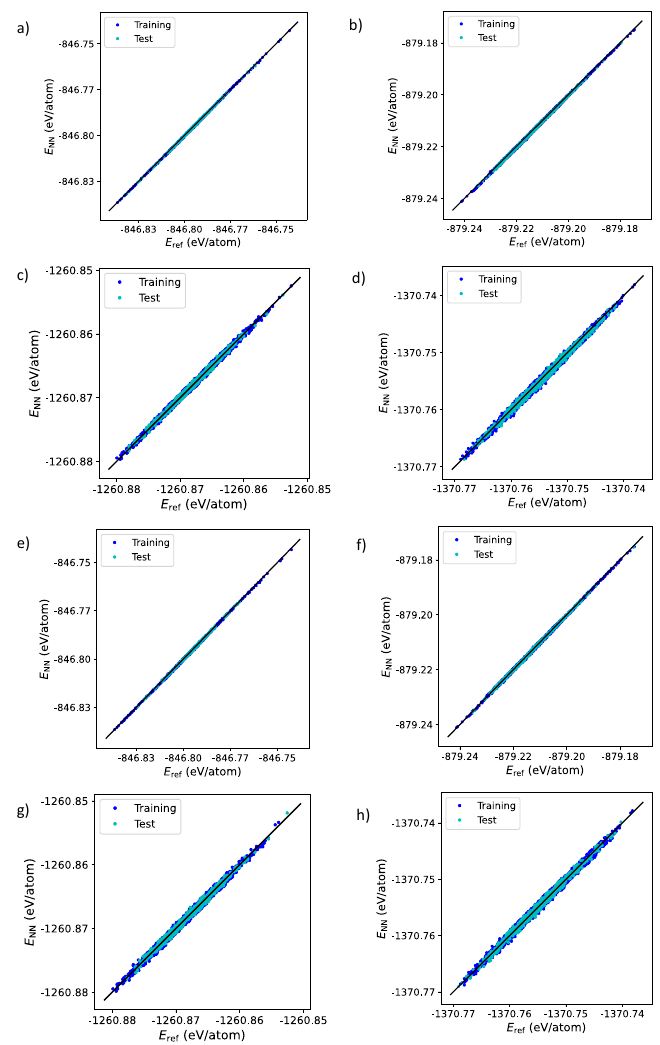}
    \caption{Correlation of the 4G-HDNNP energies $E_{\mathrm{NN}}$ and the DFT energies $E_{\mathrm{ref}}$ per atom for medium-sized FeCl$_2$ structures predicted by 4G(Fe$^{2+}$) (a), medium-sized FeCl$_2$ structures predicted by 4G(Fe$^{2+}$/Fe$^{3+}$) (b), small FeCl$_2$ structures predicted by 4G(Fe$^{2+}$) (c), small FeCl$_2$ structures predicted by 4G(Fe$^{2+}$/Fe$^{3+}$) (d), medium-sized FeCl$_3$ structures predicted by 4G(Fe$^{3+}$) (e), medium-sized FeCl$_3$ structures predicted by 4G(Fe$^{2+}$/Fe$^{3+}$) (f), small FeCl$_3$ structures predicted by 4G(Fe$^{3+}$) (g) and small FeCl$_3$ structures predicted by 4G(Fe$^{2+}$/Fe$^{3+}$) (h). }
    \label{fig:energy_correlation4G}
\end{figure*}

\begin{figure*}[!]
    \centering
    \includegraphics[width=0.95\textwidth]{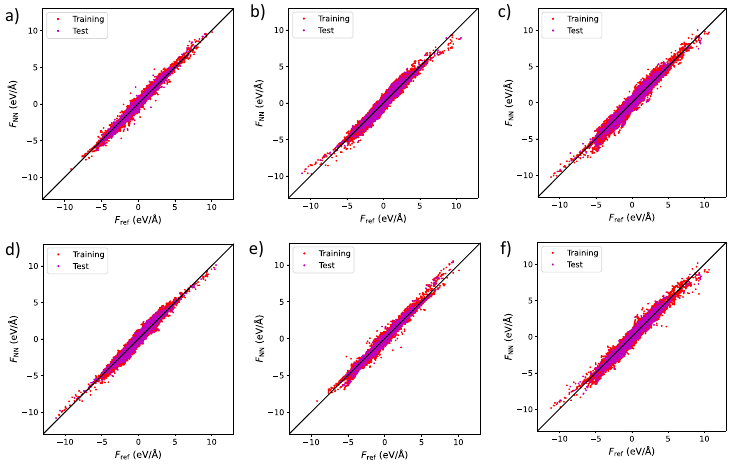}
    \caption{Correlation of the 2G-HDNNP and 4G-HDNNP atomic force components $F_{\mathrm{NN}}$ and the DFT reference atomic force components $F_{\mathrm{ref}}$ for HDNNP 2G(Fe$^{2+}$) (a), HDNNP 2G(Fe$^{3+}$) (b), HDNNP 2G(Fe$^{2+}$/Fe$^{3+}$) (c), HDNNP 4G(Fe$^{2+}$) (d), HDNNP 4G(Fe$^{3+}$) (e), and HDNNP 4G(Fe$^{2+}$/Fe$^{3+}$) (f). }
    \label{fig:force_correlation}
\end{figure*}

Figures \ref{fig:charge_correlation}, \ref{fig:energy_correlation2G}, \ref{fig:energy_correlation4G} and \ref{fig:force_correlation} present error correlations of all six HDNNPs used in this study.

\subsection{Molecular dynamics}

\begin{figure}[!]
    \centering
    \includegraphics[width=0.45\textwidth]{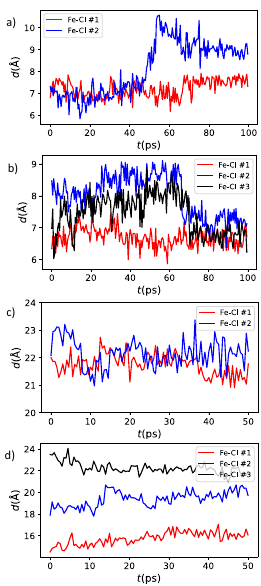}
    \caption{Fe-Cl distances in the MD simulations of FeCl$_2$ (a) and FeCl$_3$ (b) in the medium box and of FeCl$_2$ (c) and FeCl$_3$ (d) in the large box obtained with HDNNP 2G(Fe$^{2+}$/Fe$^{3+}$).
    } 
    \label{fig:2Gdistances}
\end{figure}

\begin{figure}[!]
    \centering
    \includegraphics[width=0.45\textwidth]{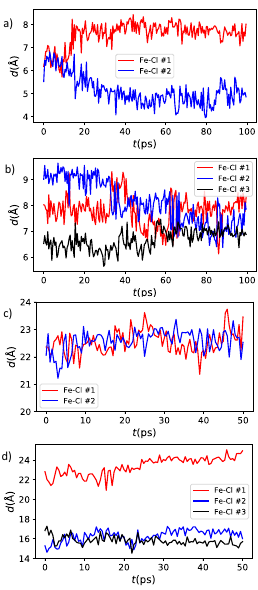}
    \caption{Fe-Cl distances in the MD simulations of FeCl$_2$ (a) and FeCl$_3$ (b) in the medium box and of FeCl$_2$ (c) and FeCl$_3$ (d) in the large box obtained with HDNNP 4G(Fe$^{2+}$/Fe$^{3+}$).
    } 
    \label{fig:4Gdistances}
\end{figure}

The Fe-Cl distances in the trajectories reported in the main text are provided in Figures \ref{fig:2Gdistances} and \ref{fig:4Gdistances}.

\bibliography{supporting.bib}